\documentclass[a4paper]{scrartcl}
\usepackage{graphicx} 
\usepackage[inkscapelatex=false]{svg}
\usepackage[style=ieee,sorting=none]{biblatex}
\usepackage{hyperref}
\usepackage{makecell}
\usepackage{authblk}
\usepackage{booktabs}
\usepackage[utf8]{inputenc}
\usepackage{lscape} 

\title{Towards Defect Phase Diagrams: From Research Data Management to Automated Workflows}
\author[1]{Khalil Rejiba}
\author[1]{Sang-Hyeok Lee}
\author[1]{Christina Gasper}
\author[1]{Martina Freund}
\author[1]{Sandra Korte-Kerzel}
\author[2]{Ulrich Kerzel}
\affil[1]{Institute of Physical Metallurgy and Materials Physics, RWTH Aachen University}
\affil[2]{Faculty of Georesources and Materials Engineering, RWTH Aachen University}

\date{October 2025}

\begin{document}

\maketitle
\small

\vspace{2.5cm}
\begin{center}
    \includegraphics[width=1.0\textwidth]{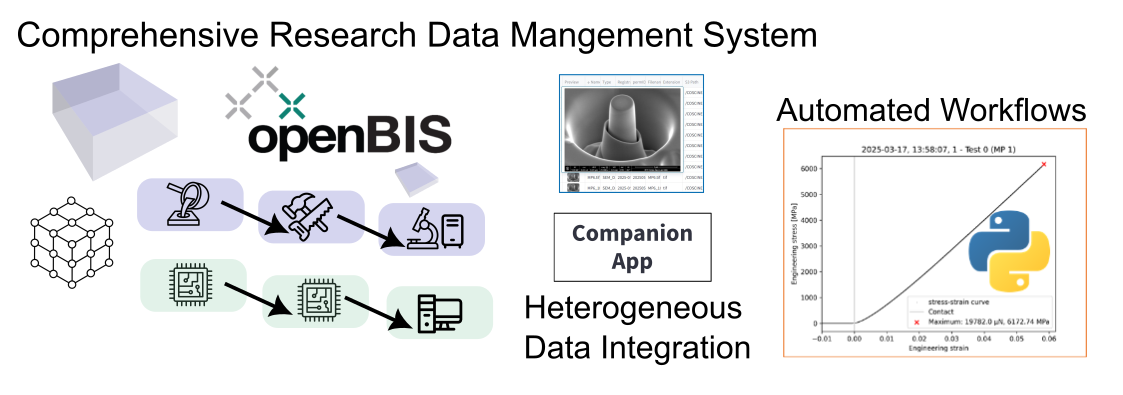}
\end{center}

\newpage
\begin{abstract}
Defect phase diagrams provide a unified description of crystal defect states for materials design and are central to the scientific objectives of the Collaborative Research Centre (CRC) 1394. Their construction requires the systematic integration of heterogeneous experimental and simulation data across research groups and locations. In this setting, research data management (RDM) is a key enabler of new scientific insight by linking distributed research activities and making complex data reproducible and reusable.

To address the challenge of heterogeneous data sources and formats, a comprehensive RDM infrastructure has been established that links experiment, data, and analysis in a seamless workflow. The system combines: (1) a joint electronic laboratory notebook and laboratory information management system, (2) easy-to-use large-object data storage, (3) automatic metadata extraction from heterogeneous and proprietary file formats, (4) interactive provenance graphs for data exploration and reuse, and (5) automated reporting and analysis workflows. The two key technological elements are the openBIS electronic laboratory notebook and laboratory information management system, and a newly developed companion application that extends openBIS with large-scale data handling, automated metadata capture, and federated access to distributed research data.

This integrated approach reduces friction in data capture and curation, enabling traceable and reusable datasets that accelerate the construction of defect phase diagrams across institutions.
\end{abstract}

\normalfont
\section{Introduction}
The field of materials science and engineering is undergoing a fundamental transformation driven by data-centric methodologies 
and digitalisation, which enable deeper insights by integrating complex, multidimensional datasets from experiments and simulations \cite{kimmig2021digital, bayerlein2024pmd}.
Crystal defects, including point, line, and planar defects, such as vacancies, solute atoms, dislocations, stacking faults, and grain boundaries, play a decisive role in determining strength, ductility, and corrosion resistance \cite{Gottstein2025}.
Within this context, the Collaborative Research Centre (CRC) 1394 is developing defect phase diagrams that are analogous to thermodynamic bulk phase diagrams and are designed to allow us to harness distinct defect states to engineer materials with tailored properties \cite{korte2022defect}. 

\begin{figure}[htp]
    \centering
    \includegraphics[width=0.75\linewidth]{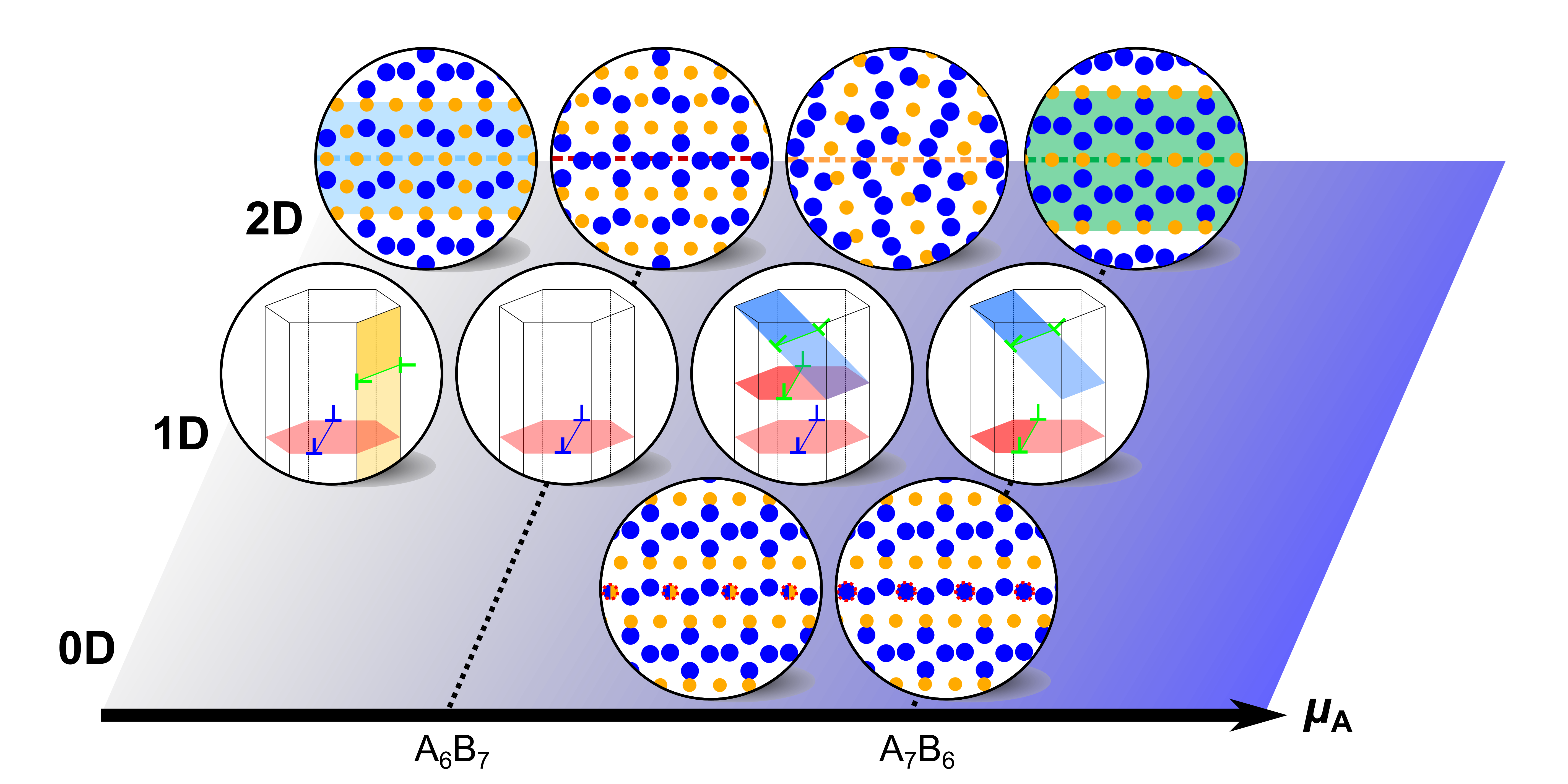}
    \caption{Example of (meta)stable defects of different dimensionalities as a function of chemical potential in $\mu$-phases with the two stoichiometric compositions $A_6 B_7$ and $A_7B_6$ as reference points. For a given type of defect, the atomic scale structure of the most stable configuration----the ``defect phase''----changes, allowing the prediction as well as tailoring of the dominant lattice defects and the resulting material properties in alloys. Reprinted from \cite{gasper2025chemicallytailoredplanardefect}, originally published under a CC BY 4.0 license.}
    \label{fig:Gasper_DPD}
\end{figure}

Realising such defect phase diagrams, which give the atomic scale composition and structure of the most stable 0D, 1D, or 2D defect states of observed lattice defects as a function of chemical potential (\autoref{fig:Gasper_DPD}), requires the comprehensive integration of experimental and computational data across diverse instruments and methods. This introduces challenges in data heterogeneity, provenance, and reproducibility: 
Each step in experimental workflows — from sample synthesis and preparation to characterisation — uses diverse instruments that generate a wide range of proprietary file formats. 
Computational methods increase the complexity further, as data from atomistic simulations to continuum modelling need to be included \cite{tehranchi2024metastable, zhou2025materials}.

In such large-scale, interdisciplinary collaborations, common practices such as local storage, transfers via physical media, or email exchanges often impede systematic documentation of data provenance, experimental details, and simulation parameters, thereby constraining collaboration.
These limitations motivate the need for a robust research data management system that integrates heterogeneous data sources, supports automated workflows, preserves end-to-end provenance, and enables reproducible analyses across distributed research groups.
These and other infrastructure-related challenges are being addressed through community initiatives such as the National Research Data Infrastructure for Materials Science (NFDI-MatWerk) \cite{eberl_2021_5082837}, which seeks shared solutions to enable research data management solutions aligned with the FAIR principles (Findable, Accessible, Interoperable, and Reusable) \cite{wilkinson2016fair}. 

In this work, we describe a solution developed and implemented by Collaborative Research Centre 1394, as a participating project in NFDI-MatWerk, to enable research data management workflows for our researchers that are efficient and easy to use in everyday laboratory or computational work. This solution allows the connection of heterogeneous data and sources crucial to our scientific goals, and results in FAIR data contributed to the scientific community.

\subsection{Requirements for Research Data Management Systems in Collaborative Environments}
As the very first step, we need to consider the requirements for a research data management system, particularly in a collaborative environment. In this setting, these systems are pivotal to the digital transformation in materials science and engineering. They help researchers manage, organise, and share data across a project’s life cycle. 
Most available systems either provide an electronic laboratory notebook (ELN) or a laboratory information management system (LIMS). However, researchers need both components to track the complete analysis chain end to end, from sample creation to final publication, including physical assets, instruments, consumables, standardised preparation steps, and experimental methods.
An electronic laboratory notebook replaces paper-based records and enables systematic documentation of experiments and results. A laboratory information management system monitors the usage of scientific instruments, tracks physical samples, records consumables such as chemicals used in experiments and preparation, and stores standardised methods. 

Building on these basic components, we identify key requirements as follows: 
\begin{itemize}
\item tracking of physical samples and performed analyses
 \begin{itemize}
    \item physical or computational samples from synthesis (or sample acquisition) to final analysis, including derived and intermediate samples,
\item processing steps applied to samples or used for their creation,
    \item tools employed, such as scientific instruments and software,
    \item data acquired from devices and generated in further analysis steps,
 \end{itemize}
\item consistent extraction of metadata for efficient and repeatable analysis workflows,
\item scalable and collaborative data storage allowing collaborators to share data,
\item intuitive graphical user interface (GUI) with extensive automation,
\item role-based access control and audit trails that satisfy institutional data governance,
\item efficient publication of relevant and curated research data according to FAIR data principles.
\end{itemize}

Several research data management systems are available as commercial and open-source solutions. 
In this work, we focus on open-source systems because they can be operated and maintained by the community and avoid recurring licensing costs that are often difficult to include in funding proposals. 

Within the open-source landscape, we evaluated Coscine \cite{lang2024rdm}, PASTA-ELN\cite{Tsybenko:972877},
eLabFTW \cite{herres2025data}, NOMAD Oasis \cite{speckhard2025big}, and openBIS \cite{barillari2016openbis}.
At the time of our decision (end of 2022), only openBIS aligned reasonably well with the combined requirements of an electronic laboratory notebook and a laboratory information management system as outlined above. 
openBIS is developed and maintained by ETH Zurich as open-source software with a large user community across the European research landscape \cite{barillari2016openbis, kuhn2022data, el2023bam, plass2023using, zaki2025self, lileikyte2025proteomic, hemani2025openbis}. 
Originally employed in biological research, it now has a substantial user base in materials science and engineering, including national research institutions such as the Federal Institute for Materials Research and Testing in Germany \cite{el2023bam} and the Swiss Federal Laboratories for Materials Science and Technology in Switzerland \cite{hemani2025openbis}. This broad community helps address the specific needs of materials science and engineering and benefits the software’s development through cross-disciplinary exchange.

\subsection*{Contribution of this Work}

This work describes how we built a comprehensive research data management system for the Collaborative Research Centre 1394, with a focus on experimental workflows.

In \autoref{sec:boundary_conditions}, we outline the boundary conditions, challenges, and opportunities in a heterogeneous, distributed research environment, including diverse instruments, proprietary file formats, and multi-institutional data governance policies.
In \autoref{openbis_mse}, we show how openBIS was customised for materials science by extending schemas for samples, instruments, protocols, and datasets; introducing controlled vocabularies and validation scripts; and implementing QR code-based sample tracking to preserve end-to-end provenance.
We illustrate these capabilities with case studies in scanning electron microscopy and micro-mechanical testing, including automated parsing of proprietary formats and efficient registration and handling of images, orientation maps, and load–displacement curves. 
In \autoref{sec:CompanionApp}, we present a companion application that connects openBIS to Coscine’s DataStorage.nrw object storage, streamlines data ingestion and retrieval, provides visual previews, and offers an interactive provenance graph for rapid discovery and reuse.
In \autoref{sec:automation}, we demonstrate automated reports and workflows that support structured documentation and standardised analyses, thereby improving reproducibility across the collaboration.

\section{Boundary Conditions, Challenges, and Opportunities} \label{sec:boundary_conditions}
\subsection{Heterogeneous Research Environment}
The scientific aims of the Collaborative Research Centre 1394 require a diverse collaboration across more than twenty research groups at four academic institutions and industrial partners, each contributing distinct methods, instruments, and information technology (IT) setups.
The research data management approach must accommodate this heterogeneity while preserving provenance and comparability across locations.

\subsubsection{Materials and Methods}
\label{sec:MaterialsMethods}
A broad range of experimental techniques generates datasets at multiple length scales that are crucial for understanding material behaviour \cite{korte2022defect}.
Bulk cast specimens often serve as starting points, while thin films may be synthesised through combinatorial deposition to explore compositional space \cite{hans2021opportunities}.
These initial samples undergo processing and are subdivided into smaller specimens to allow characterisation 
by different groups in the collaboration, each of which providing a distinct expertise, focus area, and experimental techniques \cite{Siemer:861207}.
The relationship between these specimens can be modelled using parent-child relationships in openBIS that are essential to capture the sample life cycle across these steps.

After metallographic preparation, detailed surface analysis using scanning electron microscopy (SEM) reveals microstructural features such as grain boundaries and phases.
Bulk samples are often subjected to micro-mechanical testing to probe phase-specific deformation behaviour, producing load–displacement curves that underpin subsequent mechanical analysis.

Transmission electron microscopy (TEM) is further used to characterise as-prepared, deformed, or corroded regions with high-resolution images including defects such as dislocations and phase or grain boundaries.
High resolution scanning transmission electron microscopy (HR-STEM) enables precise mapping of atomic columns and local compositional variations, for example using electron energy loss spectroscopy (EELS).
Energy-dispersive X-ray spectroscopy (EDS / EDX) and atom probe tomography (APT) provide further sources of compositional information across a sample surface (SEM and TEM) or including three-dimensional spatial locations and chemical identities of atoms (APT). In both types of electron microscopes, diffraction data is used to gather information on crystal structure and orientation with electron backscatter diffraction (EBSD) in particular providing large datasets from sample surfaces. 

\begin{figure}[htp]
    \centering
    \includegraphics[width=0.75\linewidth]{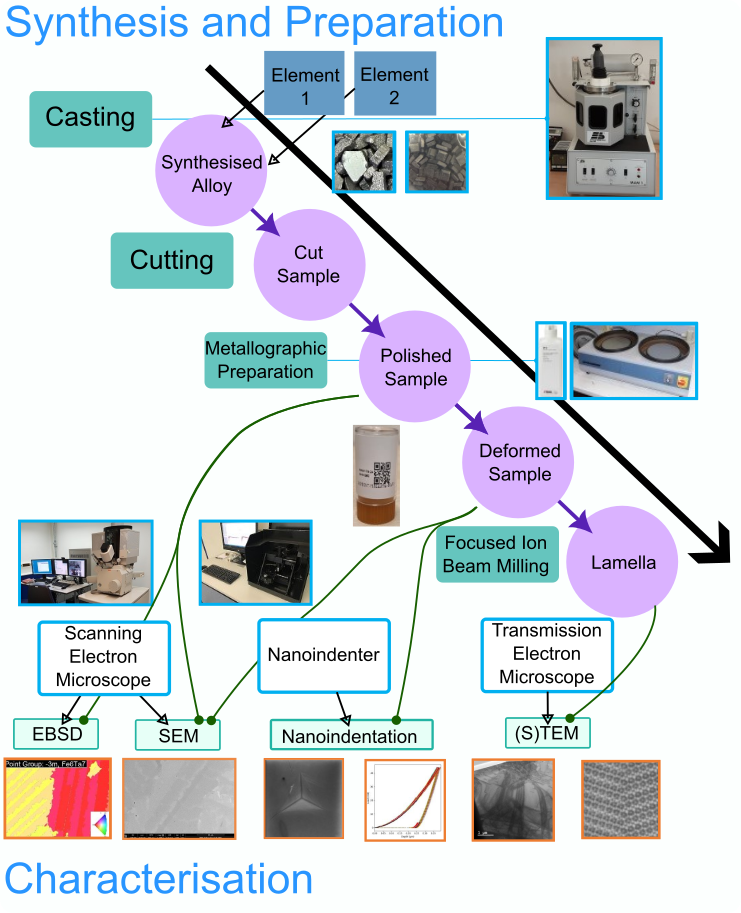}
    \caption{
    Example of a synthesis and characterisation workflow typically performed in the CRC. Devices and Consumables are indicated with a light blue border, and examples of data produced are indicated with an orange border. The thick black arrow highlights how one sample can undergo multiple processing and characterisation steps.
    }
    \label{fig:typ-workflow}
\end{figure}

\autoref{fig:typ-workflow} visualises a typical sample life cycle in the Collaborative Research Centre 1394, showing how a single specimen progresses through synthesis, preparation, characterisation, and data generation across multiple groups and instruments.

The thick black arrow highlights how one sample can undergo multiple processing and characterisation steps and how data are produced at multiple points, as described in more detail below:
Work often begins with bulk cast specimens as starting material, or with thin films synthesised by combinatorial deposition (sample creation). 
After synthesis or acquisition, samples undergo metallographic preparation (such as grinding and polishing) and are subdivided into smaller specimens (by cutting or ion beam milling) so different groups can perform complementary analyses in parallel (sample preparation).
In this example, a polished sample is initially characterised at its surface. Scanning electron microscopy with EBSD reveals microstructural features including grains and phases along with their orientations, providing foundational images and diffraction data for subsequent interpretation (surface and microstructural analysis).
Nanomechanical testing probes phase, composition, or orientation-specific deformation behaviour and produces load–displacement curves that serve as a basis for analyses of the mechanical behaviour and properties (nanomechanical analysis).
Using focused ion beam milling (FIB) and orientation information from the previous EBSD analysis, electron-transparent lamellae can be extracted from the performed nano-indents for conventional or high-resolution (S)TEM analysis of the induced dislocations, stacking faults or other defects within deformed regions.

In this way, information is gathered across a range of specimens of decreasing size from an original casting, different methods are used jointly to extract data on structure and properties with correlated conditions under which these observations took place. Together, this ensemble of physical specimens, acquired data, and their connections, allow us to extract relevant information for the creation of a defect phase diagram and mechanism/property maps -- if we can record, store, and access all other experimental and computational data sources and methods (density functional theory, molecular dynamics, mathematical image analysis, machine learning,...) relating to the same set of thermodynamic equilibria.

The diagram in \autoref{fig:typ-workflow} distinguishes equipment and consumables with light blue borders (for example, microscopes, preparation equipment, abrasives, and lubricants), emphasising the importance of recording device settings, attachments, and materials used to ensure reproducibility and comparability across institutions. 
Orange-bordered elements denote examples of data products generated along the workflow, such as micrographs, orientation maps, spectra, and mechanical response curves.

The devices and experimental setups used by each team are distributed across institutions. Each group employs specialised and often highly individualised instruments for preparation and analysis.
Each device has unique attachments, capabilities, limitations, calibration states, and setup requirements, often coupled with specific software versions and environmental conditions.
Central documentation of these details ensures that variations between experiments driven by device settings and configurations can be tracked and accounted for, thereby enhancing reproducibility and comparability across institutions.

\subsubsection{Data and Information Technology} \label{data_it}
One of the main challenges in multi-institutional collaboration is that research groups often employ similar experimental and computational methods but use different instruments and software environments.
This diversity in hardware, analysis tools, and computational infrastructure results in non-uniform or non-interoperable data formats, metadata structures, and access protocols, which complicate seamless data exchange and joint analysis. 
Institution-specific data governance policies further constrain data management and sharing across sites.

To address these challenges, the research data management system must provide a unified framework that accommodates all researchers regardless of location or local regulations.
It should integrate all data formats used by the collaboration and handle them consistently, shifting the burden of data availability and exchange from individual researchers to the system. 
Ideally, researchers interact with the system through common interfaces and workflows to contribute and access data, 
independent of data origin or local computing environment.

\subsection{Large-Scale Data Storage}
One of the main challenges of establishing an integrated laboratory information  management system is to incorporate and handle all data recorded from the scientific instruments, make them accessible for the subsequent data analysis, and track the provenance of the data and results throughout the life cycle of the entire process.

This challenge has two major aspects: the technical setup and user access.
On the technical side, the storage system must accommodate multiple terabytes of data. 
While data from nano-indentation may be relatively small, techniques such as scanning electron microscopy, electron backscatter diffraction, transmission electron microscopy, and atom probe tomography produce substantial datasets; 
for example, electron backscatter diffraction measurements can reach hundreds of gigabytes per file.

The storage system should be resilient against failure, provide backups, and ideally be geo-redundant to safeguard against catastrophic data loss.
On the researcher side, the data storage system needs to be available across a wide range of devices, from the scientific instruments from which the data are uploaded, to the individual analysis workstations. The data storage system needs to be accessible for all researchers, regardless of their location and their IT setup as discussed in \autoref{data_it}. 

While openBIS allows the integration of a large storage system as part of the instance setup, the requirements are best met by Coscine \cite{lang2024rdm}. 
The Coscine project was established in a collaboration of IT centres from multiple universities in the state of North Rhine-Westphalia (NRW), Germany. DataStorage.nrw (previously RDS-S3) is one of the provided solutions and is built on S3 storage technology \cite{bocchi2014cloud}. 
This service offers storage capacities of up to 125 terabytes per project and per application, backups, and geo-redundant storage.

While openBIS does not natively support the S3 storage technology, it does support the concept of ``Linked Data''. 
In this model, files remain in an external system, and openBIS stores links and metadata describing the content and where to find it.
Because this mechanism was developed in a different context, further integration is required to use it effectively in our setup; we address this in the companion application described in \autoref{sec:CompanionApp_DataHandling}.

\section{Customising openBIS for Materials Science} \label{openbis_mse}

\subsection{Initial Setup and Custom Objects}
\label{sec:openBIS_Setup}
\begin{figure}
    \centering
    \includegraphics[width=1\linewidth]{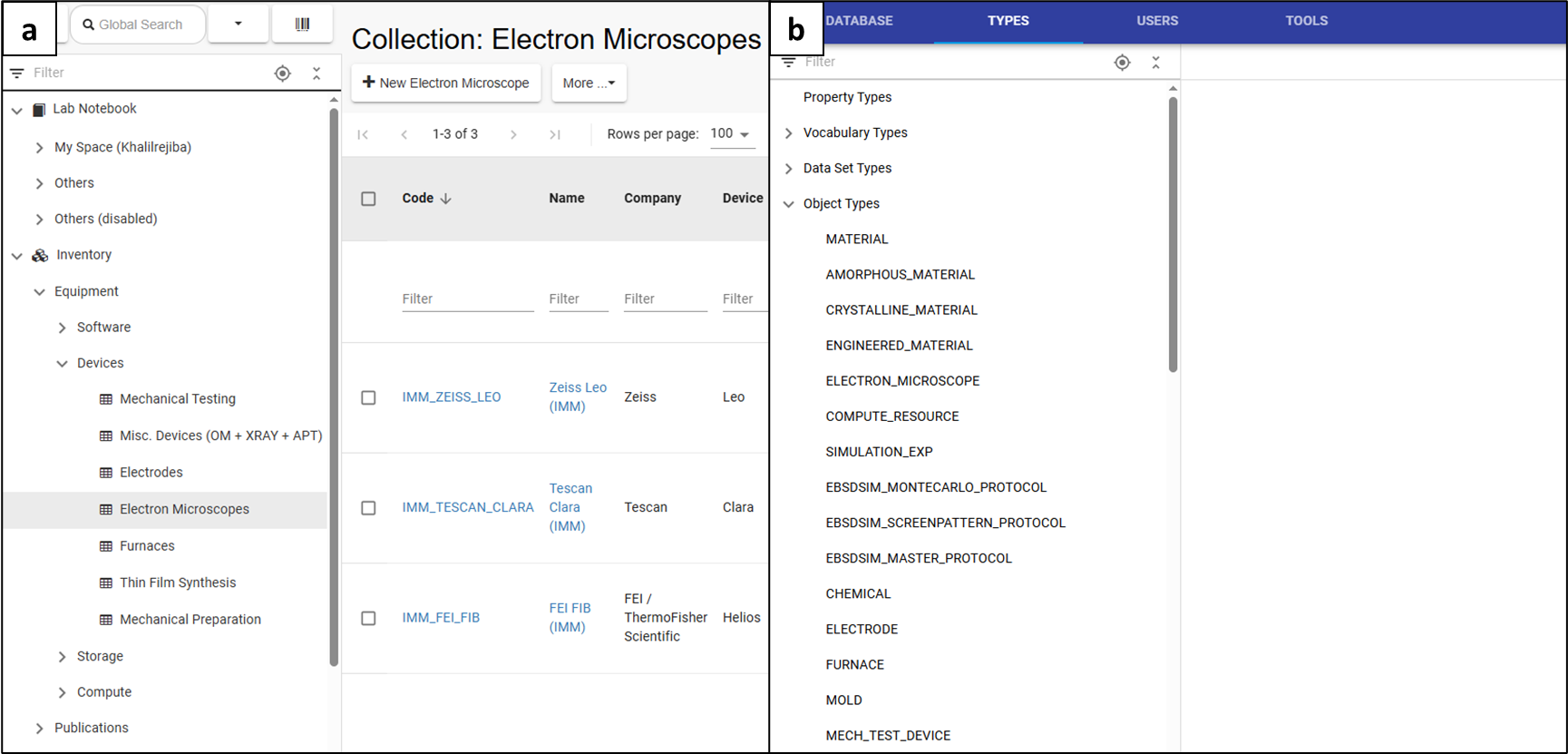}
    \caption{Tailored openBIS setup for materials science. (a): ELN-LIMS graphical interface highlighting the (reusable) inventory of instruments, methods, and consumables visible to the researcher. (b): Examples of relevant object types displayed in the Admin GUI.}
    \label{fig:openbis_configured}
\end{figure}
The default installation of openBIS provides a wide range of tools to describe and represent all stages of a scientific experiment, from sample synthesis to the experiment itself, and the subsequent analysis. 
Before the system can be used in a specific domain, it must be tailored to the scientific discipline, including descriptions of instruments, software, materials, consumables, and standardised methods used across experiments and simulations. 
In openBIS terminology, physical and digital entities—such as samples, instruments, chemicals, and protocols—are represented as ``Objects'', whereas ``Datasets'' are the associated files.

As a first step, all relevant object and dataset types are defined together with the fields to be completed by the researchers or populated through automated processes \cite{rejiba_2025_15731469}. 
Creating and extending these schemas is typically iterative because vendors differ in the metadata they embed, even for the same method, and terminology is not harmonised throughout the community.
This process is time-intensive in a collaborative setting, where a joint set of definitions and terminology must be agreed upon as an important initial step, but it also provides an opportunity to engage not only internally but also with the larger materials science and engineering community to harmonise terminology and to prepare links to emerging knowledge graphs and ontologies.

Once the object types are defined, the openBIS instance is populated with instruments, chemicals, consumables, software, and protocols.
This enables comprehensive tracking of the experiment or simulation life cycle, establishing connections between steps and the instruments and methods used at each stage.

openBIS supports validation scripts that can perform initial checks on user-entered data and controlled vocabularies that present predefined terms. The latter improves the quality of the data by establishing a common terminology across all researchers.
\autoref{fig:openbis_configured} shows a tailored openBIS instance: the user-facing electronic laboratory notebook and laboratory information management system interface (a) and the administrator view of object types (b). 
Researchers can reuse the inventory of instruments, methods, and consumables by linking the relevant parts to their electronic laboratory notebook entries.
This creates a comprehensive record of the work and reduces manual data entry, while preserving provenance and enabling automation based on structured metadata.

\subsection{Sample Tracking} \label{sec:openBIS_SampleTracking}

In the physical world, experiments are centred around the physical specimen that is used and the instruments or methods it is subjected to. 
Similarly, from a research data management perspective, samples must also be addressed in two ways. 
First, the physical sample needs to be tracked and linked to the steps, data, and results produced during the experiment (this section). 
Second, the sample must be represented in the electronic laboratory notebook and laboratory information management system and linked to its physical counterpart (\autoref{sec:openBIS_ELN}).

In openBIS, each object is associated with an immutable identifier (\textit{permId}) that can be rendered as a QR code label. By creating an object for each sample and attaching the QR code label to the physical specimen, researchers link the physical item to its electronic record and can retrieve provenance via mobile scanning in the openBIS interface. 
This enables tracking as samples move between groups and supports later archiving.

A key part of the scientific work in the Collaborative Research Centre 1394 is to compare and connect experimental results with simulation studies. 
To allow consistent handling of both sample types while preserving details unique to experiments or simulations, one digital representation is created for both experimental and computational samples. 

\begin{figure}[htp]
    \centering
    \includegraphics[width=1\linewidth]{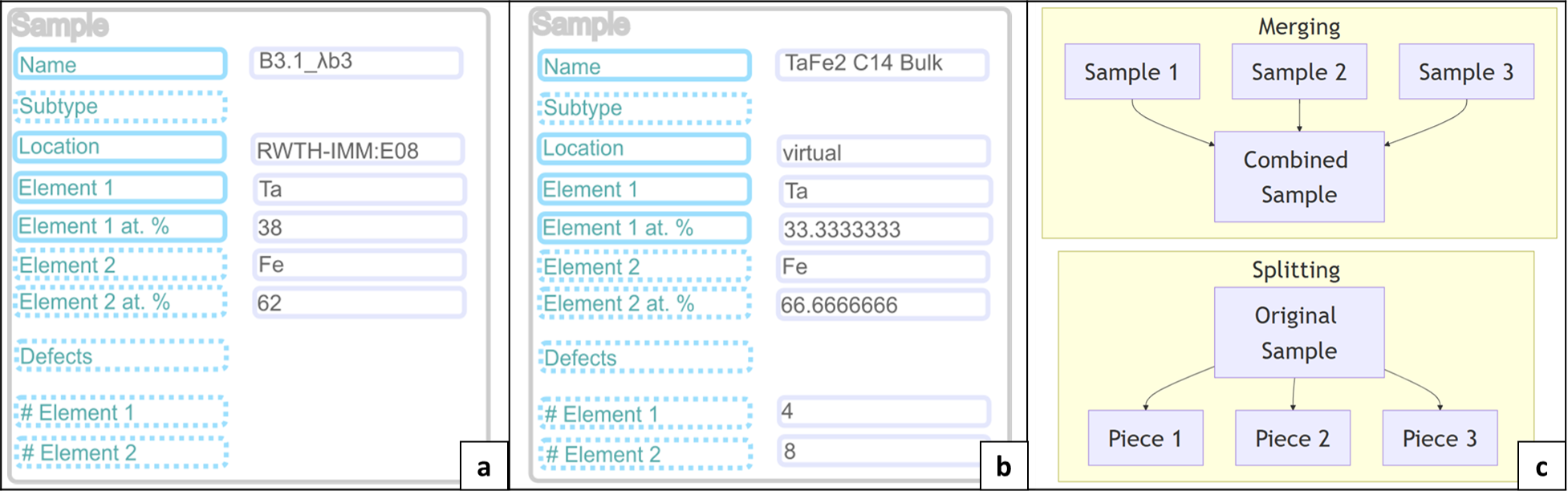}
    \caption{The sample \textit{Object Type}. Comparison of an experimental sample (a) and a computational sample (b), the dashed line highlights optional fields. Examples of how multiple samples can be linked through parent-child relationships (c).}
    \label{fig:sample-compare}
\end{figure}

We define this \textit{Object Type} as follows:
The schema includes compulsory fields, such as dimensions and location,  so that the server rejects entries when critical information is missing. 
Composition is stored as separate fields, one per chemical element, and validation scripts perform automated data quality checks; for example, the total composition must sum to 100 \%. 
A controlled vocabulary categorises experimental samples as shown in \autoref{tab:sample-classification}, presented to researchers as a drop-down list to enforce a common terminology across the collaboration. 
Because defects are central to the research questions, researchers can annotate samples with defect tags. Examples of metadata stored in the ``Sample'' \textit{Object Type} are included in \autoref{fig:sample-compare}.

\begin{table}[h]
    \centering
    \caption{\label{tab:sample-classification}A non-exhaustive list of terms used to categorise experimental samples}
    \begin{tabular}{@{}ll@{}}
        \toprule
        Sample Type  & Description \\ \midrule
        Rod          & Long and narrow sample produced by extrusion \\
        Sheet        & Flat and thin solid sample \\
        Bulk         & Generic macroscopic specimen \\
        APT Tip      & Sharpened needle for Atom Probe Tomography \\
        TEM Lamella  & Ultra-thin slice for Transmission Electron Microscopy \\
        Thin Film    & Thin coating on a substrate \\ 
        Micro Pillar & Small vertical structure fabricated to investigate mechanical properties \\
        \bottomrule
    \end{tabular}
\end{table}

In order to reflect the ``history'' of a sample in terms of sample synthesis, acquisition, and preparation, openBIS allows parent–child relationships between objects, thereby modelling a sequence of events. 
Here, we can make use of this functionality to store information such as ``Sample A underwent operation B before operation C'' and we encourage researchers to create as many intermediate samples as needed to mirror the physical reality of the sample. 

Documentation within the openBIS instance guides researchers in assigning parents and children consistently across groups, see \autoref{fig:pillar_doc} for further details. 
By preserving these relationships, researchers can systematically reconstruct a sample's entire experimental history and correlate experimental outputs with corresponding simulations.

\autoref{fig:sample-compare} illustrates the concept of the sample \textit{Object Type} developed in this work, using data from \cite{gasper2025mechanical, 
gasper2025chemicallytailoredplanardefect} as representative examples. The left panel shows the digital representation of an experimental specimen, whose electronic record links to a QR code that can be attached to the corresponding physical sample. 
The middle panel depicts the equivalent digital object used to represent a simulation input derived from the 
same material system.
As data generation within the Collaborative Research Centre 1394 continues to accelerate, we have focused on building and validating the underlying research data management framework in parallel. 
Using representative, well-characterised datasets from \cite{gasper2025mechanical, gasper2025chemicallytailoredplanardefect}, we developed and tested the workflows and digital representations of samples in openBIS. 
This proactive approach ensures that the infrastructure is fully prepared to handle and interlink forthcoming experimental and simulation data as the collaborative research advances.
The right part of the figure illustrates how parent-child relationships can be used to describe how samples are made. Multiple samples can be merged together, for example, during welding. More commonly, a large sample is cut into several pieces to allow different characterisations.

\subsection{Electronic Laboratory Notebook} \label{sec:openBIS_ELN}

Electronic laboratory notebooks enable researchers to document their work in a detailed and organised manner, replacing physical paper notebooks to track experiments across synthesis, preparation, characterisation, and analysis.
In the following, we use micro-mechanical testing combined with scanning electron microscopy to illustrate how openBIS, 
tailored to materials science, supports end-to-end documentation and provenance.

Researchers start by creating a new ``Experiment'' in openBIS, either in their personal space, or, ideally, in a shared space. 
The shared space allows multiple researchers to work in the same environment and remains accessible even as collaborators join or leave the project.

Samples used in the experiment can be sourced externally or synthesised in-house, and added either to the same space where the experiment entry is created or to a general inventory available to all researchers, which additionally stores devices and consumables.
In the case of the Collaborative Research Centre 1394, we have created a common area for all the samples from the collaboration. This enables use of a single openBIS instance across multiple research groups while maintaining access to common sample repositories.

Researchers can then add all relevant details regarding sample preparation, for example consumables and instruments used in grinding and polishing, as well as a session at the electron microscope to record the data in this specific example.

Because many details are defined during the initial setup of the openBIS installation, researchers need to provide relatively few new entries. They reuse pre-defined entities for consumables, devices, and standardised preparation methods, link them to their experiments, and record only parameters that they vary, along with environmental conditions such as temperature and humidity when suitable sensors are not available. This also includes records that are not contained in the files produced by the scientific instruments but that are essential for the subsequent analysis. 

Using controlled vocabularies as drop-down menus, as well as spreadsheets in the electronic laboratory notebook, the researchers can be supported in adding the required information manually, providing a structured approach to guide them through the process. This can then be used to develop automated workflows discussed in \autoref{sec:AutomatedAnalysis}.

\begin{figure}
    \centering
    \includegraphics[width=0.5\textwidth]{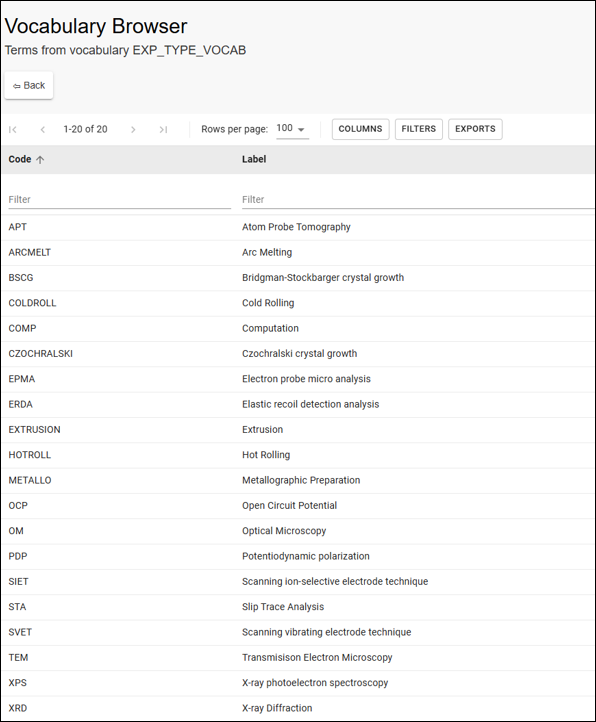}
    \caption{A view of the Vocabulary Browser in the ELN-LIMS GUI showcasing categories of experimental techniques.}
    \label{fig:experiment_type}
\end{figure}

\autoref{fig:experiment_type} shows a controlled vocabulary for experimental techniques used throughout the collaboration.
Using a common terminology establishes a shared language across groups, improves data quality by preventing divergent acronyms or abbreviations, and provides a robust foundation for automation.

Even when researchers need to develop a custom approach, they benefit from the pre-defined entities. 
They can copy an existing object and modify only the fields that require change; for example, starting from a metallographic preparation protocol and extending the etching time by editing the timing field. 
The electronic laboratory notebook entry also contains records of the data captured from scientific instruments described in more detail in \autoref{sec:data_metadata}.

\begin{figure}[htp]
    \centering
    \includegraphics[width=0.75\textwidth]{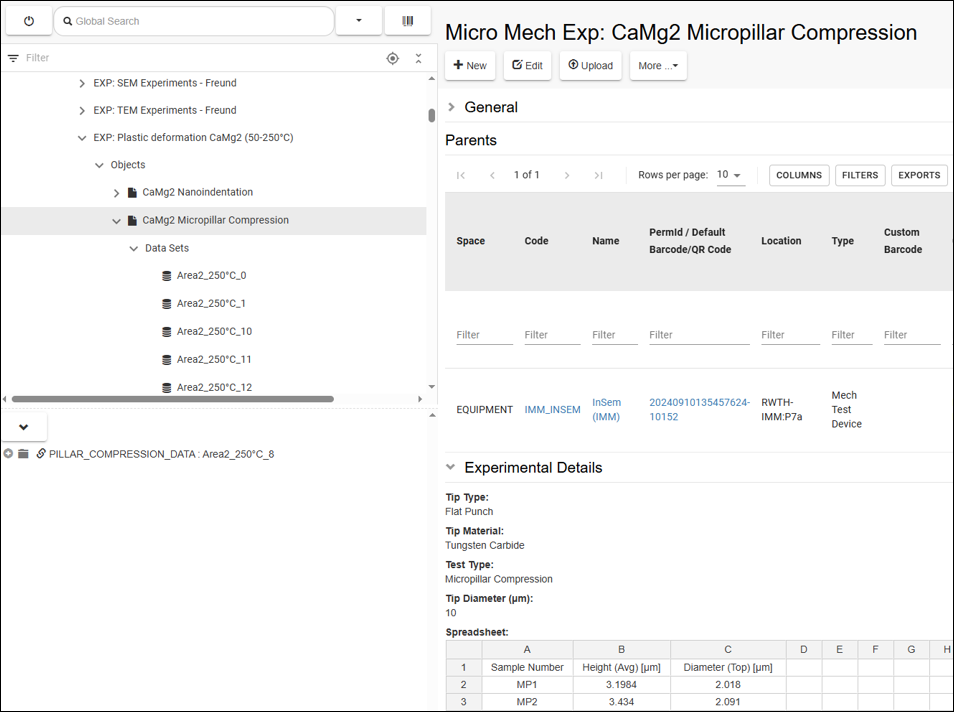}
    \caption{An example of how a micro-mechanical testing step can be represented in openBIS. Details about the micro-pillar compression experiment are recorded in a standardised manner.}
    \label{fig:eln_mp_comp}
\end{figure}

\autoref{fig:eln_mp_comp} illustrates the final entry in the electronic laboratory notebook for a micro-pillar compression test.
The left-hand side of the figure shows the navigation bar, which contains multiple micro-mechanical tests below the ``Objects'' folder. 
The right-hand side shows the experiment entry with the object type and device details used in this test. 
A spreadsheet captures key information such as micro-pillar dimensions, enabling automated workflows as highlighted in \autoref{sec:AutomatedAnalysis}.

\begin{figure}
    \centering
    \includegraphics[width=1\linewidth]{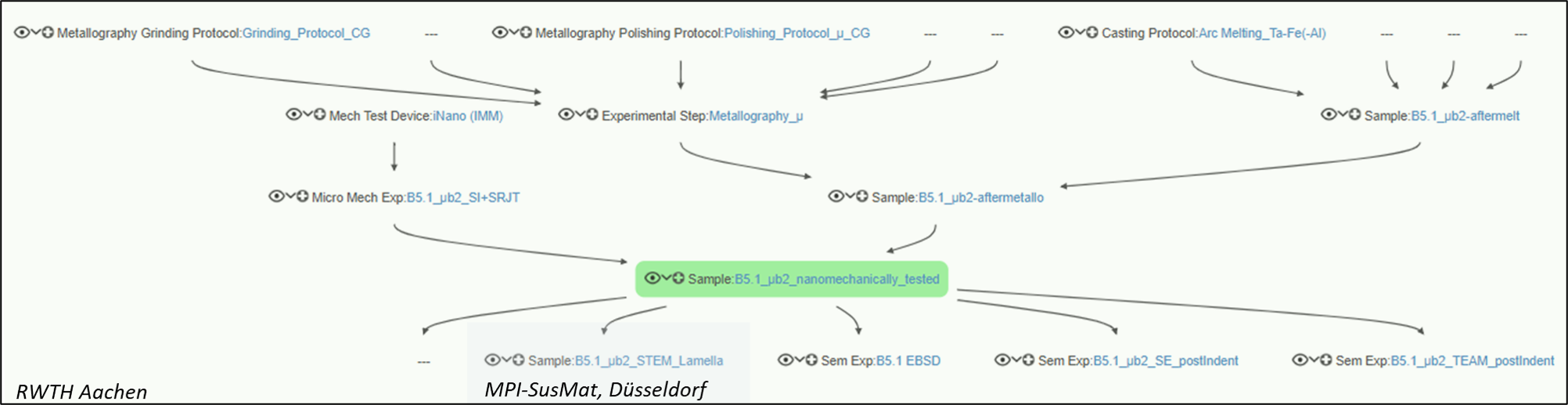}
    \caption{Graphical representation of the provenance and history of a sample that underwent nano-indentation and was characterised by SEM and EBSD. The sample is then transferred across laboratories and undergoes further characterisation The sample from which the query starts is highlighted in green.}
    \label{fig:parent_child_example}
\end{figure}

\begin{figure}
    \centering
    \includegraphics[width=1\linewidth]{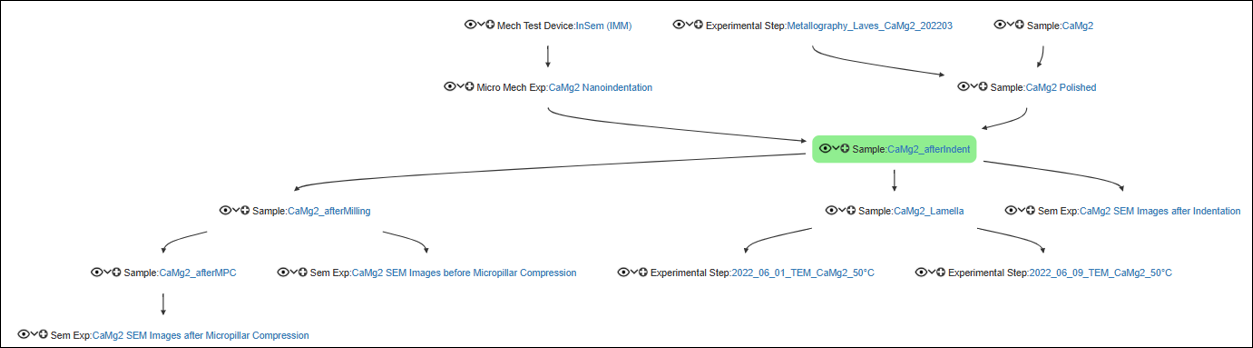}
    \caption{Graphical representation of the provenance and history of a sample that underwent both nano-indentation and micro-pillar compression and characterised by SEM and conventional TEM. The sample from which the query starts is highlighted in green.}
    \label{fig:parent_child_second_example}
\end{figure}

Since all parts of the analysis life cycle can be captured in a similar way, the relations between these entries, 
as well as the instruments, methods, consumables used across the chain can be tracked to establish full provenance. 
\autoref{fig:parent_child_example} shows a hierarchical graph of linked samples, processes, and devices and represents a concrete example from the workflow shown earlier in \autoref{fig:typ-workflow}. 
Each node of the graph can be explored by clicking on the respective element in the graphical interface.
The figure is built around the sample highlighted in green in the second-to-last row of the graph.
Moving upwards, researchers can follow the sample’s provenance from synthesis to the experiments where it was used.
``Protocols'' represent standardised scientific procedures such as arc melting, metallographic polishing or grinding. 
``Devices'' represent scientific instruments, together with their description, such as the nano-indentation device.
``Exp'' or ``Experimental Step'' represent entries in the electronic laboratory notebook by the researchers. 
The experimental step is a generic template, whereas ``Exp'' has been specifically tailored for micro-mechanical testing or scanning electron microscopy to capture relevant information.
Parts of the graph have been omitted and replaced by ``$\ldots$'' to keep the figure readable.
In \autoref{fig:parent_child_second_example}, we highlight a different sample from an earlier study \cite{freund2021plastic}. Moving downwards from the indented sample highlighted in green, researchers can follow subsequent analysis and processing steps. On the left-hand side, micro-pillars are created in a bulk sample using ion milling and are imaged in a scanning electron microscope before and after compression. On the right-hand side, we highlight the fabrication of an electron-transparent lamella, from which data is collected in a transmission electron microscope on two separate instances.

\subsection{Handling Data and Metadata}
\label{sec:data_metadata}
Data captured from scientific instruments or simulation outputs typically contain rich metadata that describe instrument settings during acquisition, software configurations, sample identifiers, and contextual information such as environmental conditions. 
Rather than asking the researchers to add these metadata manually, the data-handling process should include automated parsing to extract relevant metadata and register them in the electronic laboratory notebook for consistent, reproducible analysis.

There are two main challenges in metadata extraction: 
First, materials science and engineering data are heterogeneous, spanning many methods and instruments, and vendors seldom provide standardised, open file formats.
As a result, each vendor typically uses proprietary, often binary formats that must be decoded before metadata can be extracted; some vendors publish specifications, while others do not, requiring reverse engineering.
For each instrument and each software version, the community must determine whether the format is known and supported, whether existing tools can read the files and extract metadata, or whether new software needs to be developed. 
Additionally, vendors employ different terminology for similar concepts, which requires frequent schema alignment and revision of object definitions in openBIS when new systems are integrated.
This challenge affects the academic community as a whole and is best addressed through coordinated efforts such as the National Research Data Infrastructure for Materials Science (NFDI-MatWerk).
Starting from the initial metadata schema developed earlier in the Collaborative Research Centre \cite{Siemer:861207}, definitions are improved and expanded as more data formats become available, which are published online \cite{rejiba_2025_15731469}.

The second challenge concerns technical integration when files are registered in openBIS.
openBIS provides ``dropboxes'' that assume that data are stored locally within the server 
and can be copied to dedicated folders on the local filesystem.
This mechanism relies on Jython \cite{juneau2010definitive} which bridges Java (in which openBIS is developed) and Python. 
However, Jython has not been updated since Python 2.7 and does not allow integration of external libraries. 
Consequently, the stack provided for metadata extraction is outdated, and community tools  for file parsing and analysis cannot be integrated through this route \cite{johnstone2020density, hakon_wiik_anes_2025_14995525, francisco_de_la_pena_2025_14956374}. 
Although the openBIS development team has recognised this limitation, a workable replacement is not available in the near term.
As a result, the existing mechanisms do not support metadata handling for cloud-based object storage, which is required by the distributed nature of Collaborative Research Centre 1394.
We, therefore, implement metadata extraction outside the legacy stack, via a Python workflow in a companion application described in \autoref{sec:CompanionApp} below.

\section{Companion Application for openBIS}
\label{sec:CompanionApp}

In order to overcome the challenges that remain after tailoring openBIS for materials science and engineering, we have developed a companion application that integrates the following functionality:
\begin{itemize}
    \item Data upload and retrieval from cloud-based object store (S3).
    \item Registration of externally stored files as linked data in openBIS.
    \item Extraction of metadata from proprietary formats and registration in openBIS.
\end{itemize}
The companion application is built with the Streamlit web development framework \cite{streamlit2025} and can be hosted on a dedicated server or local machines. It operates separately from openBIS and communicates with the server through the Python application programming interface, pyBIS. 
While this requires the reseacher to interact with two graphical interfaces, it enables integration of interactive visualisations, community-developed tools, and project-specific features, as well as tailoring the functionality to the specific needs of the researchers of the collaboration. 
In the following, we describe the key aspects of the companion application in more detail.

\subsection{Adding Data and Metadata to ELN Entries with the Companion App}
\label{sec:CompanionApp_Interaction}

\begin{figure}[htp]
    \centering
    \includegraphics[width=15cm]{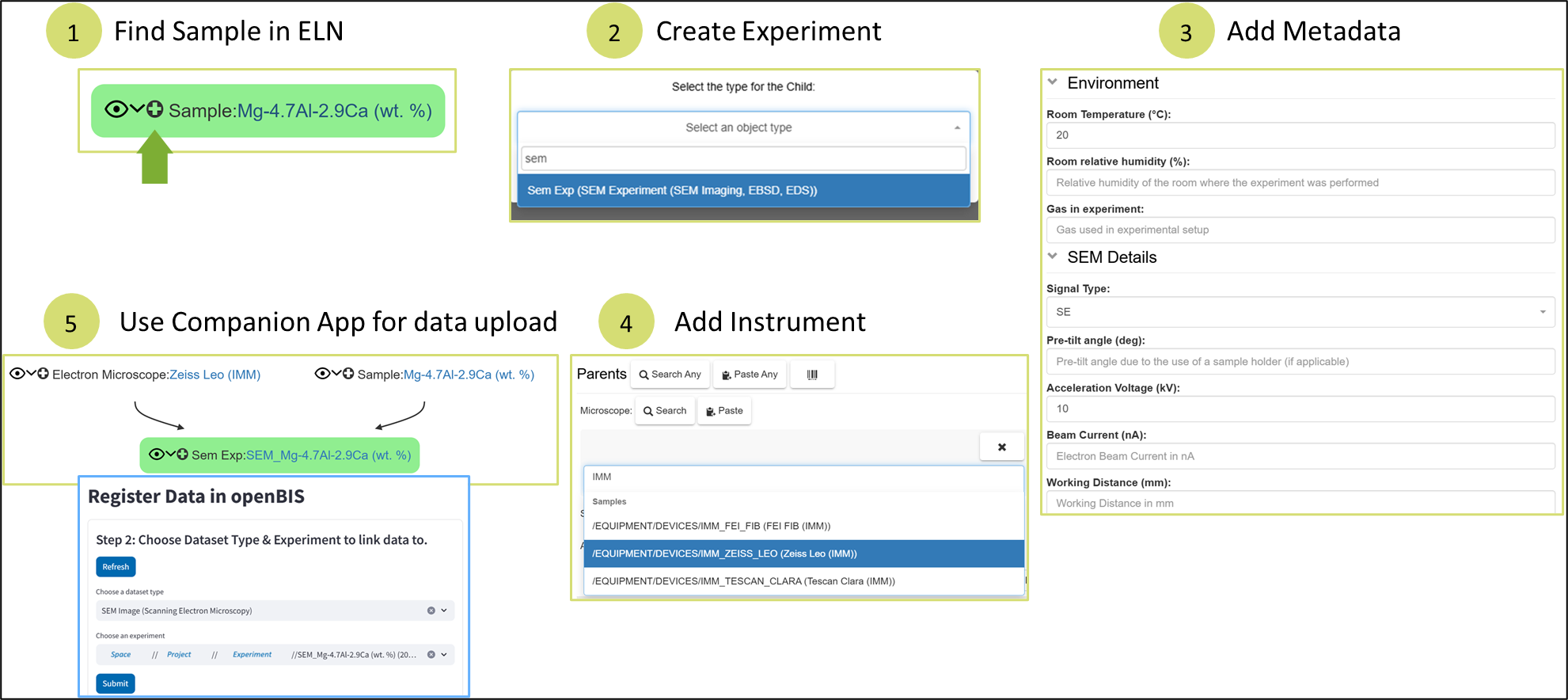}
    \caption{Example workflow of how researchers use the electronic laboratory notebook with the companion application to add data and metadata to ELN entries.}
    \label{fig:ux}
\end{figure}

The companion application allows researchers to add relevant data and metadata in the electronic laboratory notebook. 
\autoref{fig:ux} summarises the overall process, and \autoref{sec:CompanionApp_DataHandling} below describes data handling and metadata extraction in more detail.

Researchers start from an existing experimental sample in the openBIS web interface either by navigating the sidebar or by scanning the physical specimen’s QR code label (step 1).
In step 2, they create an entry in the electronic laboratory notebook, to store, for example, images acquired by a scanning electron microscope or measurements using electron backscatter diffraction, energy-dispersive X-ray spectroscopy, or electron channelling contrast imaging.  
Using parent-child relationships in openBIS, this notebook entry is linked to the sample, establishing provenance.
In step 3, researchers complete the notebook entry. For scanning electron microscopy, a customised template with a moderate number of fields reduces manual input because most fields can be extracted automatically from the uploaded files (see \autoref{sec:CompanionApp_DataHandling}).
In step 4, the scientific instrument is linked (for example, a specific scanning electron microscope model), which automatically adds instrument details to the experiment without requiring manual intervention.
Finally, in step 5, the researchers use the companion application to upload the data to the corresponding entry in the electronic laboratory notebook, specifying the \textit{Dataset Type}, which defines the fields that are filled automatically during metadata extraction.

\subsection{Data Handling and Automatic Metadata Extraction} 
\label{sec:CompanionApp_DataHandling}
\begin{figure}
    \centering
    \includegraphics[width=1\linewidth]{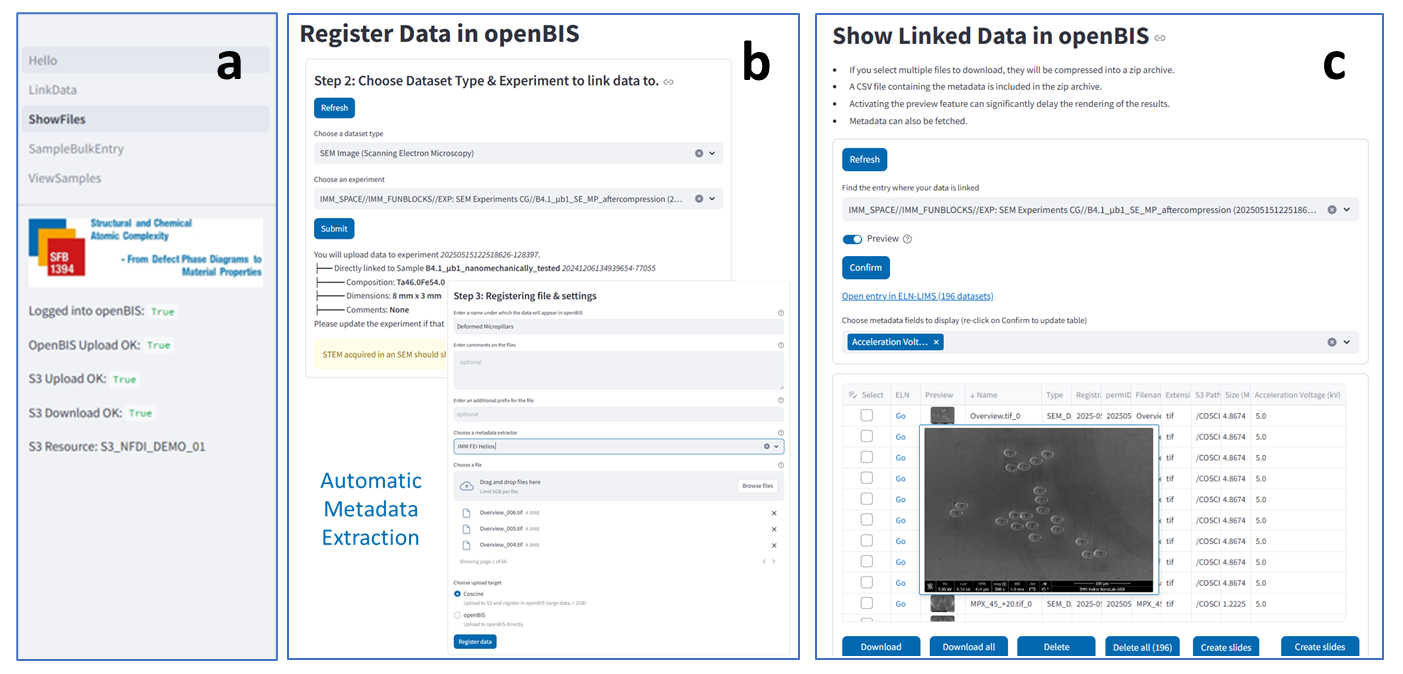}
    \caption{An overview of the companion application. A sidebar (a) enables the researcher to navigate between pages and actions. One page (b) is dedicated to registering data and another (c) is used for data retrieval.}
    \label{fig:companion_app_overview}
\end{figure}

\autoref{fig:companion_app_overview} illustrates the process from the researcher's perspective: 
First, the researchers need to authenticate both with openBIS, and the cloud object storage to obtain write access. 
This also gives the researchers access to the relevant parts in openBIS inside the companion application and allows them to upload files to the external storage. 
The companion application then retrieves the electronic laboratory notebook entries that the researcher is authorised to access, and the researcher selects the experiment to which the data should be linked.

\begin{figure}[htp]
    \centering
    \includegraphics[width=1\linewidth]{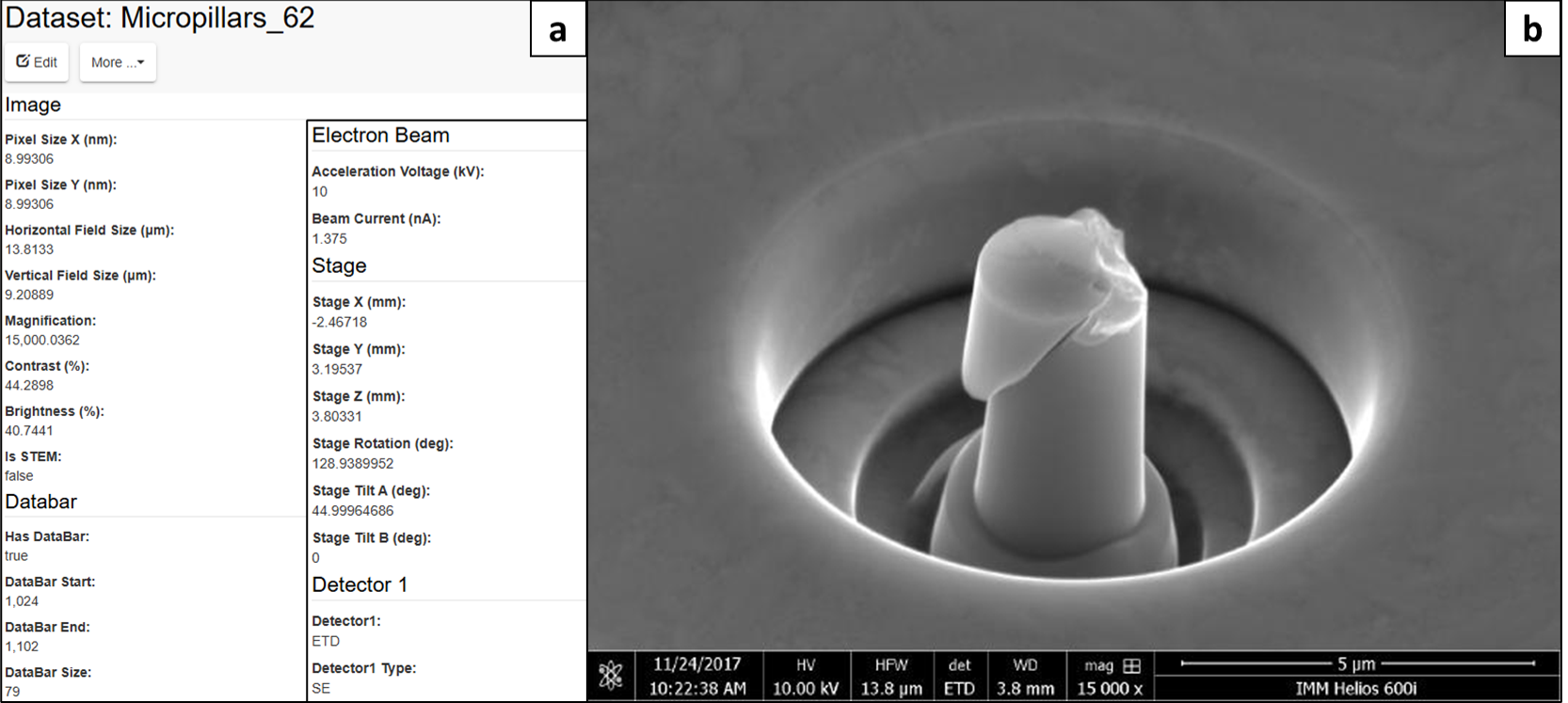}
    \caption{Extracted SEM metadata view in the ELN-LIMS GUI (a) alongside corresponding image of a deformed micro-pillar (b)}
    \label{fig:sem_metadata}
\end{figure}

Because a file type alone does not uniquely identify the scientific instrument or the software used to generate it, researchers must select the appropriate metadata parser from a list of available options before uploading data. 
Scientific instruments typically record data in a variety of vendor-specific file formats, each with its own structure and conventions. 
As a result, each format requires a dedicated method to extract information, and the internal data structure often differs between vendors, even for the same instrument type.
Vendors also use different terminology for equivalent physical quantities. 
In scanning electron microscopy, for example, \autoref{tab:sem_metadata} shows that vendor A labels the acceleration voltage as ``EHT,'' vendor B as ``HV,'' and vendor C as ``Beam.HV.'' 
To ensure consistent metadata representation within the research data management system, we first define a unified superset of terms and then map vendor-specific names to this vocabulary in the metadata parsers. 
This vocabulary can subsequently link to a formal ontology \cite{bayerlein2024natural, hofmann2025em}, which supports semantic interoperability and enables integration of the data into knowledge graphs. 
\autoref{fig:sem_metadata} illustrates how openBIS displays the extracted metadata together with the corresponding micrograph, allowing researchers to access both data and metadata directly.
When no parser exists, researchers can still register and link the file; they can later enrich it with metadata once a parser becomes available.

\begin{figure}[htp]
    \centering
    \includegraphics[width=0.75\textwidth]{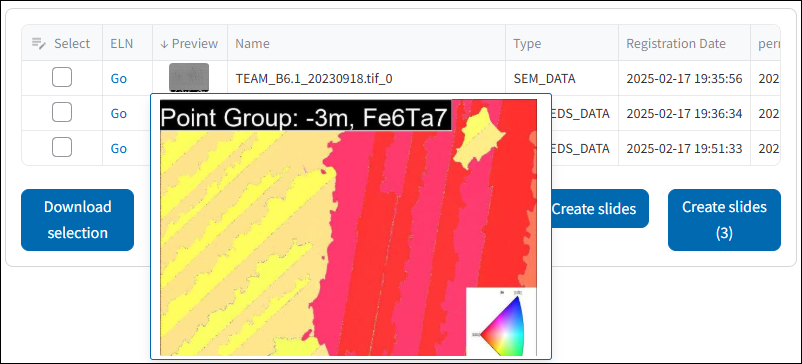}
    \caption{Example of EBSD data visualised in the companion application.
    The orientation map is extracted from the EBSD data and stored as a preview image. 
    Then, clicking on this thumbnail, a larger version, shown here in the screenshot, is presented in the companion application. Researchers can then quickly verify that the files 
    hold the relevant information before downloading and analysing them in more detail. 
    Data taken from \cite{gasper2024preparation}.}
    \label{fig:data_visualisation}
\end{figure}

These challenges also imply that researchers typically cannot open files easily and verify that they have accessed the right data they are looking for.
The companion application addresses this limitation with an integrated preview function similar to modern file browsers. 
It automatically generates thumbnails for image data and tailored previews for non-image data where possible. 
As shown in \autoref{fig:data_visualisation}, the system can extract an orientation map from electron backscatter diffraction data and store it in the cloud as a preview image. 
This functionality allows the researcher to confirm file content before downloading and processing. 
Researchers can select and download multiple files simultaneously. 
The companion application also includes a ``Create slides'' feature that lets researchers compile selected images into a slide deck for sharing. Each slide contains one image and a corresponding table of metadata, as shown in \autoref{fig:slides}.

\subsection{Finding Samples through an Extended Hierarchical Graph}

The openBIS system itself already provides hierarchical graphs as discussed earlier and illustrated in \autoref{fig:parent_child_example} and \autoref{fig:parent_child_second_example}.
However, the native graph’s appearance depends on where the researcher opens it in the interface, and its interaction model is relatively static. Although researchers can expand linked entities and view metadata, expanded nodes quickly consume screen space and make navigation cumbersome.

\begin{figure}
    \centering
    \includegraphics[width=0.75\textwidth]{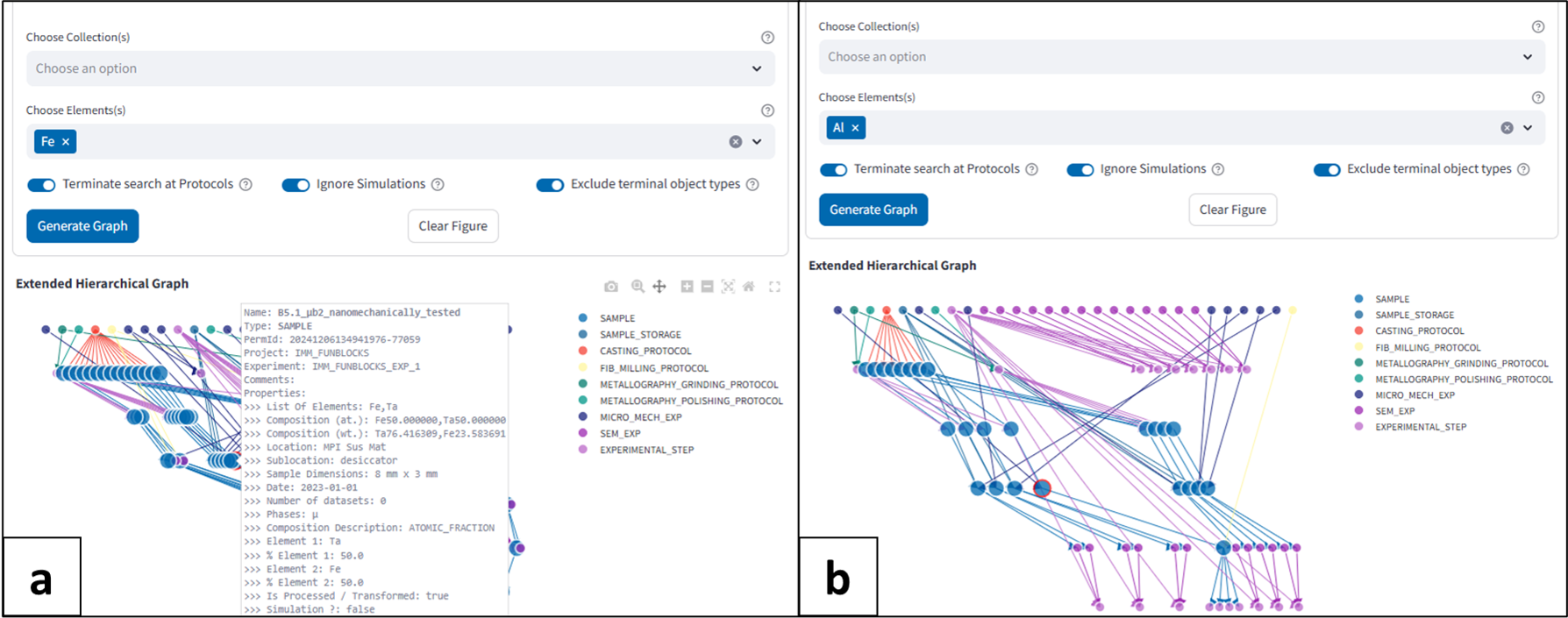}
    \caption{Extended hierarchical graph viewed in the Companion App. The graph shows all samples, methods, and ELN entries for all samples containing iron (a) and aluminium (b). The additional metadata that is accessible by hovering the mouse cursor over the respective element is additionally illustrated in (a). This figure uses the same experimental sample as \autoref{fig:parent_child_example}.}
    \label{fig:companion_app_hierarchical_graph}
\end{figure}

To improve sample discovery and provenance exploration, the companion application includes an extended, interactive hierarchical graph as shown in \autoref{fig:companion_app_hierarchical_graph}.
The graph uses colours to distinguish entity types such as samples, protocols, experiments, instruments, and it presents a consistent layout regardless of the entry point. 
Starting from material elements, researchers can filter to find samples that contain a specific chemical element, then traverse linked protocols, instruments, experiment entries, and associated datasets. 
Researchers can hover over any node to reveal key metadata in a tooltip, such as composition, dimensions, 
instrument model, acquisition date, and dataset counts, and click on nodes to open the corresponding record. 

\section{Workflows and Automation} \label{sec:automation}
So far, combining openBIS as an electronic laboratory notebook and a laboratory information  management system with the companion application developed in this work enables researchers to create a single source of truth.
The system tracks provenance from sample creation through analysis and records all relevant data associated with the research.
Once the research data management system is established in the everyday practice of the researchers, its functionality can be extended beyond a documentation and archival tool: because data and their provenance are accessible within one environment, automated workflows can be built on top of the established structures.

For example, routine analysis steps can be integrated into the data registration or executed as scheduled tasks.
This benefits researchers in several ways. 
First, standardised data-quality checks and baseline analyses are executed automatically, ensuring consistency across researchers and reducing manual effort. 
Second, a common software environment delivers identical results for identical inputs, improving reproducibility. 
Third, automation alleviates the burden of running routine analyses manually, allowing researchers to focus on specialised tasks aligned with their scientific objectives.

In the following, we provide two examples that illustrate the benefits of this approach: automated reports and automated standard analysis.

\subsection{Automated reports}
\label{sec:AutomatedReport}
 
A key part in making research reproducible is to establish the provenance of not only the scientific results and the samples, 
instruments, and software used, but also of the steps required to obtain those results.
For example, before a metallic sample can be used in micro-mechanical testing, electron microscopy, or other analysis, it must be prepared metallographically as appropriate for the analysis.
The detailed steps can be stored as a ``standard operating procedure'' (SOP) or a ``protocol'' in openBIS and linked to the relevant entry in the electronic laboratory notebook.

This has two main advantages: 
First, the method is documented already at the beginning of the analysis, and no detail is lost.
Second, procedures can be documented not only in an individual researcher’s space but also in group spaces or the global inventory, which reduces duplication and enables reuse. Technicians, student helpers, and other researchers can then draw on this shared toolbox of methods for their work.

\begin{figure}[htp]
    \centering
    \includegraphics[width=10cm]{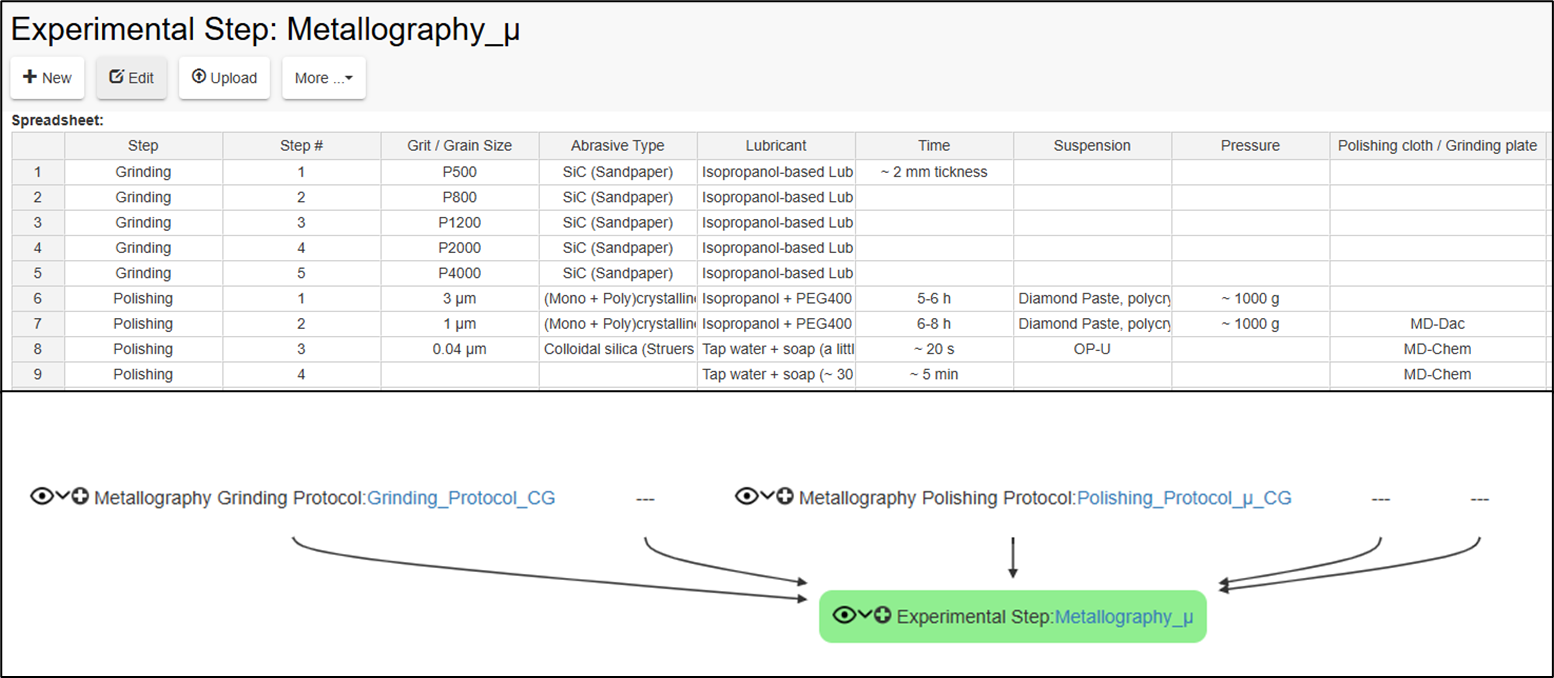}
    \caption{Automatically generated summary report of metallographic preparation steps in the ELN-LIMS graphical interface. 
    All details are extracted from data entered by the researcher, showing the abrasive materials and lubricants used in the preparation of the sample. Data taken from \cite{gasper2024preparation}.
    }
    \label{fig:metallo}
\end{figure}

To make this more accessible to the researchers, we have implemented an automated report that summarises the procedure in a table (\autoref{fig:metallo}). 
Researchers can quickly verify that all steps are appropriate for the planned work, and the report provides a convenient template for documenting methods in subsequent publications.

\subsection{Automated Analysis in micro-mechanical testing} \label{sec:AutomatedAnalysis}
Micro-pillar compression testing \cite{uchic2004sample} is a micro-mechanical testing technique in which focused ion beam fabricated pillars, typically a few micrometres in diameter, are compressed using a nano-indenter to probe the plastic deformation behaviour of small volumes of material. It enables quantitative investigation of size effects, slip mechanisms, and local microstructural phenomena under well-controlled, near uniaxial stress conditions, bridging the gap between nano-indentation \cite{Oliver1992AnIT} and standardised bulk mechanical testing.
\begin{figure}[htp]
    \centering
    \includegraphics[width=10cm]{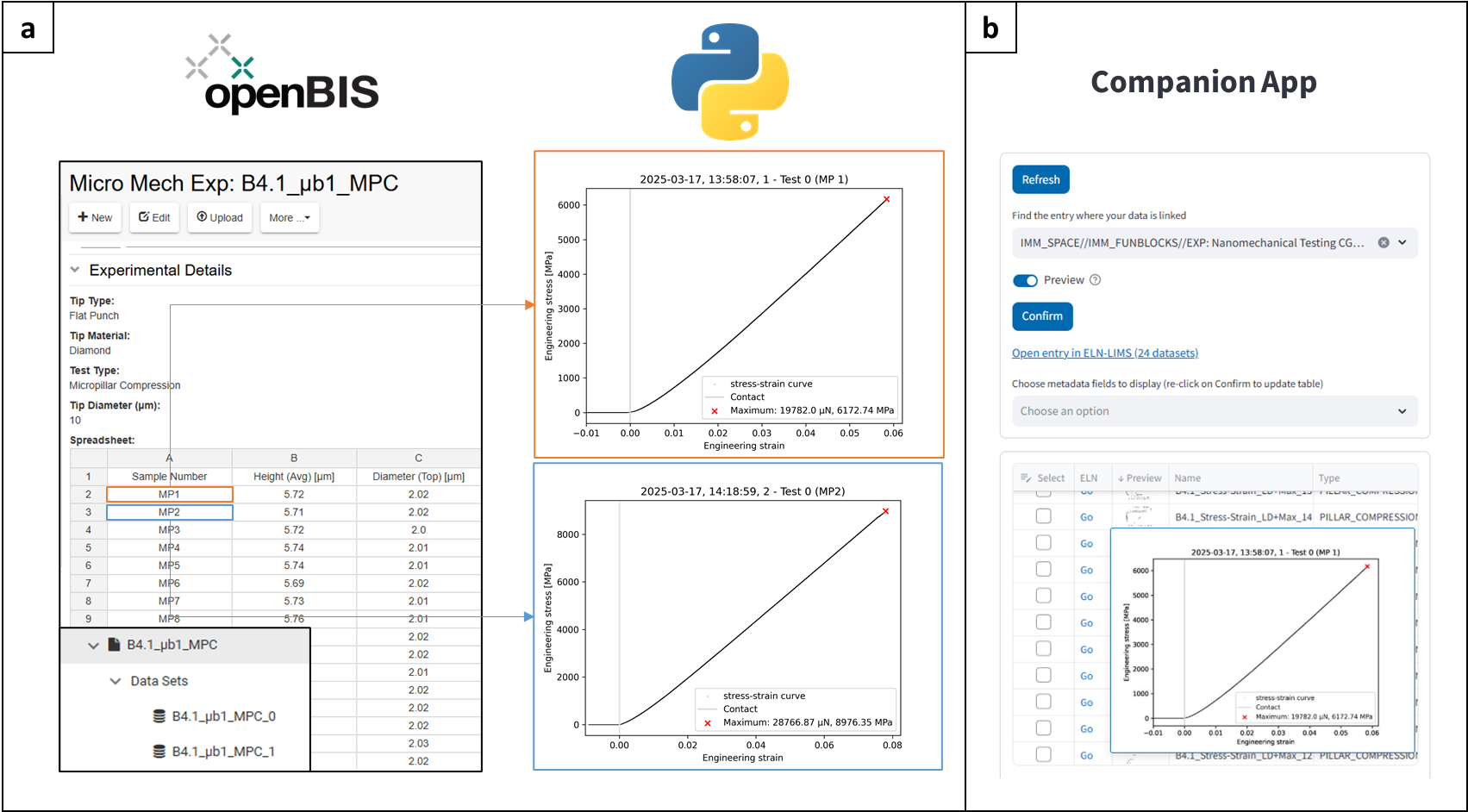}
    \caption{Automated analysis in openBIS for the example of micro-pillar compression tests. An ELN entry for micro-mechanical testing (``Micro Mech Exp''), containing details for multiple micro-pillars milled in the sample (a).
    Using two examples (MP1, orange, and MP2, blue), the stress-strain curves can be automatically derived from the data uploaded by the researcher and shown in the bottom left corner. The axes for each stress-strain curve have been rescaled using the dimensions of the micro-pillar. The figures are uploaded back to the ELN entry and can be viewed in the companion application (b).}
    \label{fig:stress-strain}
\end{figure}
A standard baseline analysis is the construction of the stress–strain curve, which shows the sample’s response under mechanical loading. 
Because this curve should be derived as a first step in further analysis, we implemented an automated workflow (\autoref{fig:stress-strain}).
The researcher creates an entry in the electronic laboratory notebook using a customised ``Micro Mech Exp'' template and fills in the relevant details as defined by the template. As shown in \autoref{fig:stress-strain} a, multiple micro-pillars milled from the same specimen
are documented, and the researcher uploads the raw load–displacement data as described in \autoref{sec:CompanionApp_DataHandling}.

A set of Python scripts, executed as a scheduled job on a dedicated server, 
retrieves the relevant data and metadata from the object storage and openBIS for each ``Micro Mech Exp'' instance. 
The workflow converts load–displacement to stress–strain using the recorded geometry,
and generates the stress–strain curves.
The figures are then uploaded as datasets to the same entry in the electronic laboratory notebook and become available in the companion application 
(\autoref{fig:stress-strain} b). 
These scripts provide a starting point for researchers to perform more specific analyses in Jupyter notebooks, ensuring consistent, reproducible baseline analyses.

\section{Conclusion and Outlook}
Establishing a comprehensive research data management system for heterogeneous collaborations requires integrating data from a wide range of scientific instruments with proprietary formats, aligning metadata schemas across disciplines, accommodating diverse information technology setups, and linking physical entities such as samples and instruments to their digital representations. 
Within the Collaborative Research Centre 1394, combining openBIS as an electronic laboratory notebook and a laboratory information management system with a companion application addresses much of this need. 
Tailored electronic records, controlled vocabularies, validation scripts, QR-code labels, and parent–child relationships enable end-to-end provenance and customisation to match research group requirements. 
The companion application complements openBIS by providing cloud storage integration, automated extraction of metadata from uploaded files, and automated reports and workflows that support structured documentation and standardised analyses.

In future work, we will expand the set of vendor-specific file formats from which metadata are extracted automatically and integrate Jupyter Hub into the openBIS setup to allow researchers to access standardised analyses and workflows directly from a web browser, with convenient access to all relevant data and metadata for a given analysis.
We will also extend the rollout of tablets across research groups to make data entry and access more convenient when moving between different parts of the laboratories without consistent access to desktop computers. 
Finally, although openBIS already provides export functionality to repositories such as Zenodo \cite{zenodo}, we will extend this to include data stored in the cloud and additional details currently available through the companion application, ensuring that published datasets are findable, accessible, interoperable, and reusable.

\section*{Acknowledgements}
Funded by the Deutsche Forschungsgemeinschaft (DFG, German Research Foundation) as part of Collaborative Research Centre CRC 1394 - Structural and Chemical Atomic Complexity - From Defect Phase Diagrams to Material Properties (project number 409476157).\\
Funded by the Deutsche Forschungsgemeinschaft (DFG, German Research Foundation) under the National Research Data Infrastructure – NFDI 38/1 – project number 460247524.\\
The data used in this publication was managed using the research data management platform Coscine (http://doi.org/10.17616/R31NJNJZ) with storage space of the Research Data Storage (RDS) (DFG: INST222/1261-1) and DataStorage.nrw (DFG: INST222/1530-1) granted by the DFG and Ministry of
Culture and Science of the State of North Rhine-Westphalia.\\
This project has received funding from the European Research Council (ERC) under the European Union’s Horizon 2020 research and innovation programme (grant agreement No. 852096 FunBlocks and grant agreement No. 101168203 TailorPlast).\\

\section*{CRediT author statement}
Khalil Rejiba: Methodology, Software, Data Curation, Writing- Original draft. 
Sang-Hyeok Lee: Software, Writing - Review \& Editing.
Christina Gasper: Data curation.
Martina Freund: Data curation.
Sandra Korte-Kerzel: Supervision, Project administration, Funding acquisition, Writing - Review \& Editing
Ulrich Kerzel: Conceptualization, Methodology, Software, Supervision, Project administration, Funding acquisition, Writing- Original draft, Writing - Review \& Editing.

\printbibliography

\pagebreak
\appendix
\counterwithin{figure}{section}
\counterwithin{table}{section}
\section{Supplementary Material}
\subsection{Documentation}
\begin{figure}[htp]
    \centering
    \includegraphics[width=0.7\textwidth]{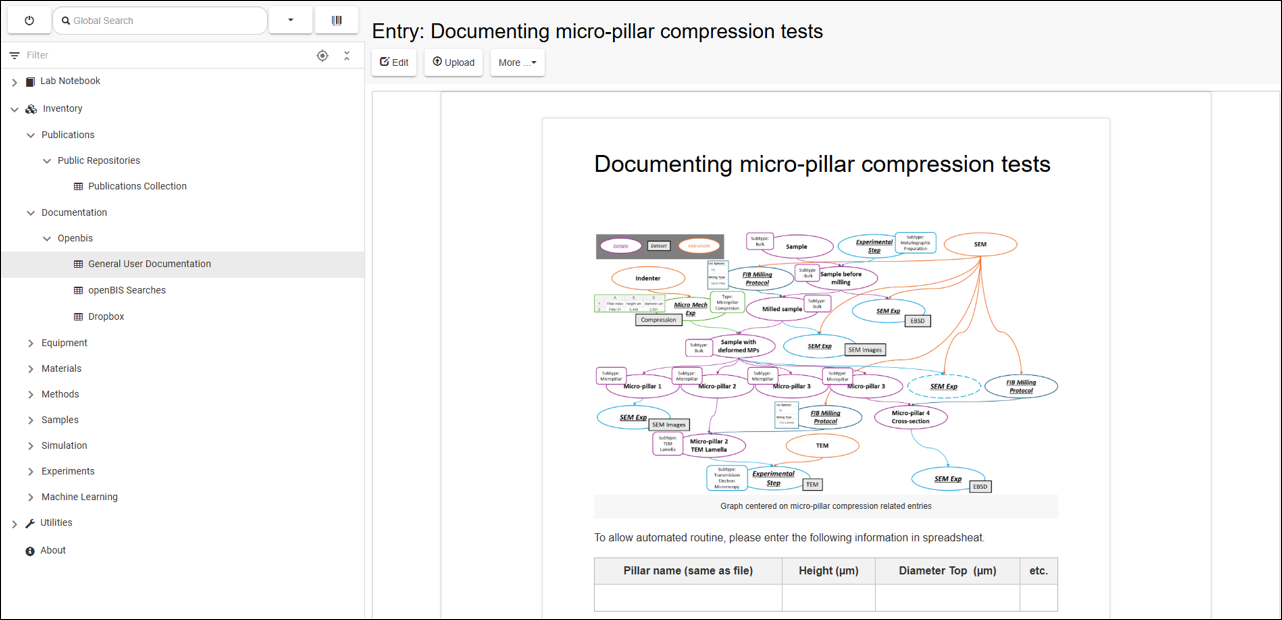}
    \caption{Screenshot showing documentation of micro-pillar compression tests stored in openBIS.}
    \label{fig:pillar_doc}
\end{figure}

\begin{figure}[htp]
    \centering
    \includegraphics[width=0.7\textwidth]{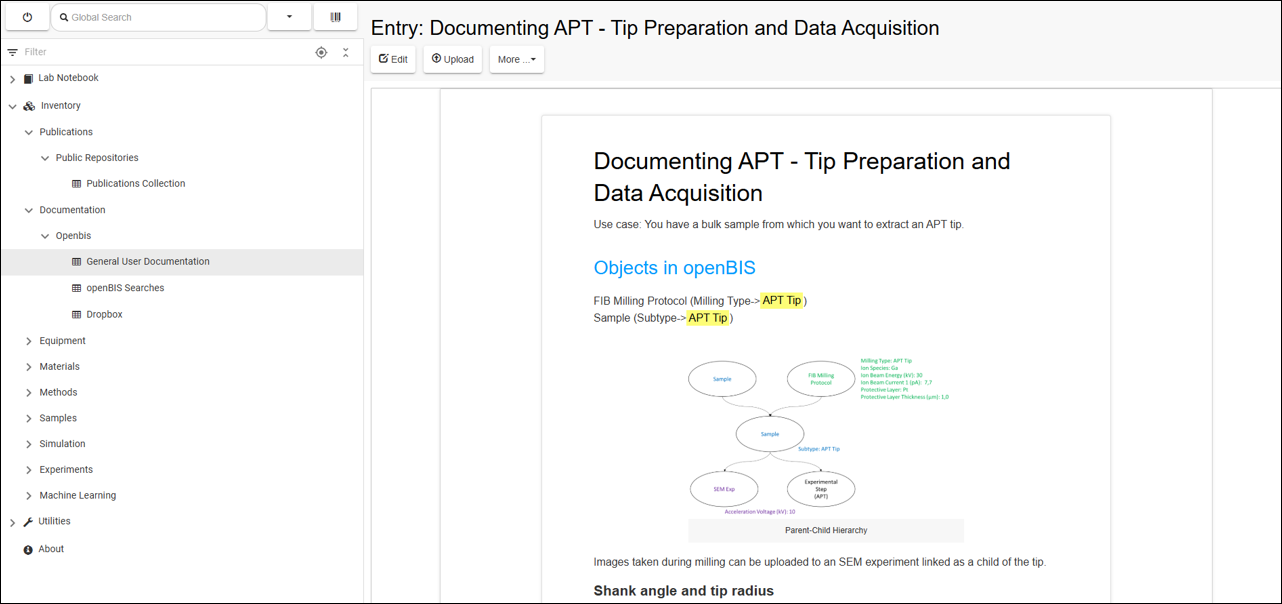}
    \caption{Screenshot showing documentation centred on atom probe tomography stored in openBIS.}
    \label{fig:apt_doc}
\end{figure}

Documentation is essential for an electronic laboratory notebook. To ensure researchers can effectively navigate the system and utilise its features, we have created documentation pages, using the ``Entry'' \textit{Object Type}, and store them in the same graphical interface as the electronic laboratory notebook. It provides clear instructions on tasks like data entry, sharing, and collaboration, preventing confusion and errors. Well-organised documentation also ensures consistency, helps maintain data integrity, and supports compliance with the FAIR principles, making it a critical resource for the researchers. In \autoref{fig:pillar_doc}, we show an example geared towards micro-pillar compression testing, where we highlight which objects need to be created and what information to be included in spreadsheets to allow automated routines. \autoref{fig:apt_doc} shows another example built around the technique of atom probe tomography.
\subsection{Metadata parsers}
Parsing files from different manufacturers is challenging due to inconsistent terminology and varying definitions for similar fields. Furthermore, metadata are not always stored consistently across manufacturers, leading to discrepancies in data representation. In \autoref{tab:sem_metadata}, we compare different fields relevant to images acquired through scanning electron microscopy across the metadata records produced by three different microscopes, each being a representative of a vendor. Fields that can be computed are indicated with ``*''. Missing values are indicated with ``x''. Links with existing ontologies are highlighted in the last column, ``?'' indicating that no one-to-one mapping has been identified.

\subsection{Companion application}
\begin{figure}[htp]
    \centering
    \includegraphics[width=0.9\textwidth]{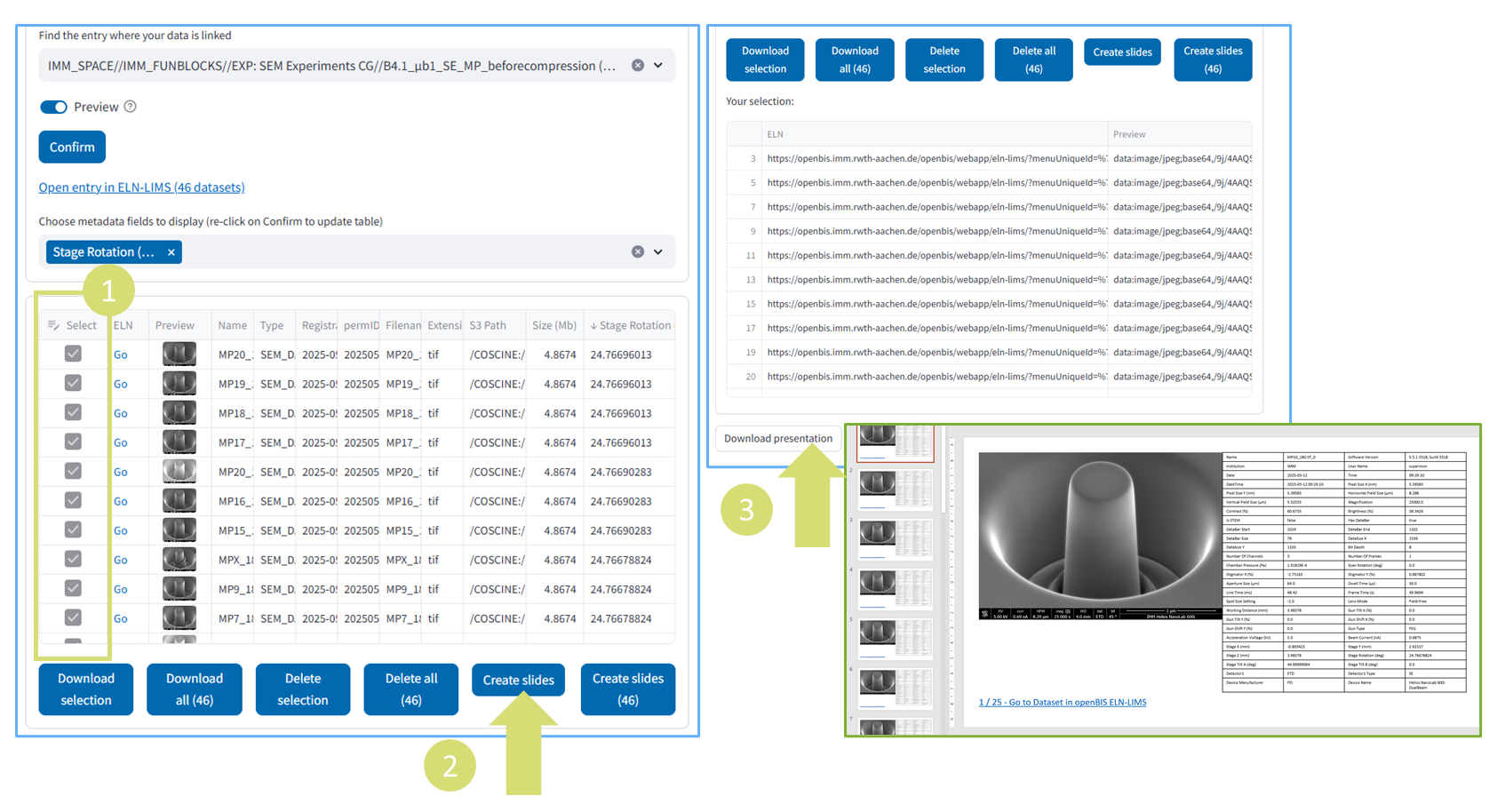}
    \caption{Example showing slide deck creation from data registered in openBIS.}
    \label{fig:slides}
\end{figure}

The need for a companion application for openBIS stems from the limitations of the graphical interface of its electronic laboratory notebook. While openBIS allows for data export, the available options are primarily geared towards archiving rather than facilitating active use in ongoing research or workflows. A dedicated companion application could bridge this gap by offering more dynamic tools for data interaction and real-time use.  In \autoref{fig:slides}, we demonstrate how the researchers can use the companion application to inspect datasets that they have already registered and create a document from selected files.  

In \autoref{fig:slides}, we show the process of creating a slide deck containing images of micro-pillars. After selecting the relevant electronic laboratory notebook entry using the companion application, the researcher can select the desired images (step 1). They can, by clicking on buttons (steps 2 and 3), create and download a set of slides containing the selected images.

\begin{landscape}
\begin{table}[h]
    \centering
    \renewcommand{\arraystretch}{1.5}
    \caption{\label{tab:sem_metadata} A non-exhaustive list of metadata that can be extracted from SEM images}
    \begin{tabular}{llllll}
        \toprule
         Field & Manufacturer A & Manufacturer B & Manufacturer C & Link to ontologies \\ \midrule
         Acceleration Voltage & EHT & HV & Beam.HV & \href{https://purls.helmholtz-metadaten.de/emg/EMG_00000004}{emg/EMG\_00000004} \\
         Dwell Time & Dwell Time & DwellTime & Scan.Dwelltime & \href{https://purls.helmholtz-metadaten.de/emg/EMG_00000015}{emg/EMG\_00000015} \\
         Stage X & Stage at X & StageX & Stage.StageX & ? \\
         Stage Y & Stage at Y & StageY & Stage.StageY & ? \\
         Stage Z & Stage at Z & StageZ & Stage.StageZ & ? \\
         Stage Rotation & Stage at R & StageRotation & Stage.StageR  & ? \\
         Working Distance & WD & WD & Stage.WorkingDistance & \href{https://purls.helmholtz-metadaten.de/emg/EMG_00000050}{emg/EMG\_00000050} \\
         Pixel Size & Pixel Size & PixelSizeX & Scan.PixelWidth & \href{https://w3id.org/pmd/mo/PixelSize}{pmd/mo/PixelSize} \\
         Emission Current & x & EmissionCurrent & EBeam.EmissionCurrent & \href{https://purls.helmholtz-metadaten.de/emg/EMG_00000025}{emg/EMG\_00000025} \\
         Beam Current & Beam Current & PredictedBeamCurrent & BeamCurrent & \href{https://purls.helmholtz-metadaten.de/emg/EMG_00000006}{emg/EMG\_00000006} \\
         Frame Time & Cycle Time & x & x & \href{https://w3id.org/pmd/mo/FrameTime}{pmd/mo/FrameTime} \\ 
         Frame Time & Line Time & x & EScan.LineTime & ? \\ 
         Magnification & Mag & Magnification & * & \href{https://w3id.org/pmd/mo/ActualMagnification}{pmd/mo/ActualMagnification} \\
         Chamber Pressure & Chamber & ChamberPressure & Vacuum.ChPressure & \href{https://w3id.org/pmd/mo/ChamberVacuum}{pmd/mo/ChamberVacuum} \\
         System Vacuum & System Vacuum & x & x & \href{https://w3id.org/pmd/mo/SystemVacuum}{pmd/mo/SystemVacuum} \\
         Gun Vacuum & Gun Vacuum & x & x & \href{https://w3id.org/pmd/mo/GunVacuum}{pmd/mo/GunVacuum} \\
         Databar Size & * & ImageStripSize & * & ? \\
        \bottomrule
    \end{tabular}
\end{table}
\end{landscape}

\end{document}